\RequirePackage[mathlines,displaymath]{lineno}
\documentclass[aps,prl,twocolumn,showpacs,amsmath,superscriptaddress,groupedaddress]{revtex4}  
\usepackage{graphicx}  
\usepackage{dcolumn}   
\usepackage{bm}        
\usepackage{amssymb}   
\usepackage[T1]{fontenc}
\usepackage[latin1]{inputenc}
\usepackage{longtable}
\usepackage{graphicx}

\usepackage{url}
\usepackage{xspace}
\usepackage{multirow}
\usepackage{units}
\usepackage[normalem]{ulem}
\usepackage{color}

\usepackage{epstopdf} 

\newcommand{\ttbar}{\ensuremath{t\bar{t}}\xspace}

\newcommand{\etmiss}{\ensuremath{E \kern-0.6em\slash_{\rm T}}\xspace}
\newcommand{\etmissx}{\ensuremath{E \kern-0.6em\slash_{\rm x}}\xspace}
\newcommand{\etmissy}{\ensuremath{E \kern-0.6em\slash_{\rm y}}\xspace}
\newcommand{\alpgen}{{\sc alpgen}\xspace}
\newcommand{\pythia}{{\sc pythia}\xspace}
\newcommand{\herwig}{{\sc herwig}\xspace}
\newcommand{\vecbos}{{\sc vecbos}\xspace}

\newcommand{\mt}{\ensuremath{m_t}\xspace}
\newcommand{\mw}{\ensuremath{M_W}\xspace}
\newcommand{\kjes}{\ensuremath{k_{\rm JES}}\xspace}
\newcommand{\pevt}{\ensuremath{P_{\rm evt}}\xspace}
\newcommand{\psig}{\ensuremath{P_{\rm sig}}\xspace}
\newcommand{\pbkg}{\ensuremath{P_{\rm bkg}}\xspace}

\newcommand{\ejets}{\ensuremath{e+{\rm jets}}\xspace}
\newcommand{\mujets}{\ensuremath{\mu+{\rm jets}}\xspace}
\newcommand{\ljets}{\ensuremath{\ell+{\rm jets}}\xspace}

\newcommand{\etal}{\textit{et al.}\xspace}

\newcommand{\GeV}{\ensuremath{\textnormal{GeV}}\xspace}

\newcommand{\dif}{\ensuremath{{\rm d}}}

\newcommand{\met}{\ensuremath{{}/\!\!\!{p}_{\rm T}}\xspace}

\newcommand{\fb}{\ensuremath{{\rm fb}^{-1}}\xspace}

\newcommand{\pt}{\ensuremath{p_{\rm T}}\xspace}

\newcommand{\x}{\ensuremath{\vec x}\xspace}
\newcommand{\y}{\ensuremath{\vec y}\xspace}

\begin{document}

\hspace{5.2in} \mbox{FERMILAB-PUB-14-123-E}

\title{Precision measurement of the top-quark mass in lepton$+$jets final states
}
%
\affiliation{LAFEX, Centro Brasileiro de Pesquisas F\'{i}sicas, Rio de Janeiro, Brazil}
\affiliation{Universidade do Estado do Rio de Janeiro, Rio de Janeiro, Brazil}
\affiliation{Universidade Federal do ABC, Santo Andr\'e, Brazil}
\affiliation{University of Science and Technology of China, Hefei, People's Republic of China}
\affiliation{Universidad de los Andes, Bogot\'a, Colombia}
\affiliation{Charles University, Faculty of Mathematics and Physics, Center for Particle Physics, Prague, Czech Republic}
\affiliation{Czech Technical University in Prague, Prague, Czech Republic}
\affiliation{Institute of Physics, Academy of Sciences of the Czech Republic, Prague, Czech Republic}
\affiliation{Universidad San Francisco de Quito, Quito, Ecuador}
\affiliation{LPC, Universit\'e Blaise Pascal, CNRS/IN2P3, Clermont, France}
\affiliation{LPSC, Universit\'e Joseph Fourier Grenoble 1, CNRS/IN2P3, Institut National Polytechnique de Grenoble, Grenoble, France}
\affiliation{CPPM, Aix-Marseille Universit\'e, CNRS/IN2P3, Marseille, France}
\affiliation{LAL, Universit\'e Paris-Sud, CNRS/IN2P3, Orsay, France}
\affiliation{LPNHE, Universit\'es Paris VI and VII, CNRS/IN2P3, Paris, France}
\affiliation{CEA, Irfu, SPP, Saclay, France}
\affiliation{IPHC, Universit\'e de Strasbourg, CNRS/IN2P3, Strasbourg, France}
\affiliation{IPNL, Universit\'e Lyon 1, CNRS/IN2P3, Villeurbanne, France and Universit\'e de Lyon, Lyon, France}
\affiliation{III. Physikalisches Institut A, RWTH Aachen University, Aachen, Germany}
\affiliation{Physikalisches Institut, Universit\"at Freiburg, Freiburg, Germany}
\affiliation{II. Physikalisches Institut, Georg-August-Universit\"at G\"ottingen, G\"ottingen, Germany}
\affiliation{Institut f\"ur Physik, Universit\"at Mainz, Mainz, Germany}
\affiliation{Ludwig-Maximilians-Universit\"at M\"unchen, M\"unchen, Germany}
\affiliation{Panjab University, Chandigarh, India}
\affiliation{Delhi University, Delhi, India}
\affiliation{Tata Institute of Fundamental Research, Mumbai, India}
\affiliation{University College Dublin, Dublin, Ireland}
\affiliation{Korea Detector Laboratory, Korea University, Seoul, Korea}
\affiliation{CINVESTAV, Mexico City, Mexico}
\affiliation{Nikhef, Science Park, Amsterdam, the Netherlands}
\affiliation{Radboud University Nijmegen, Nijmegen, the Netherlands}
\affiliation{Joint Institute for Nuclear Research, Dubna, Russia}
\affiliation{Institute for Theoretical and Experimental Physics, Moscow, Russia}
\affiliation{Moscow State University, Moscow, Russia}
\affiliation{Institute for High Energy Physics, Protvino, Russia}
\affiliation{Petersburg Nuclear Physics Institute, St. Petersburg, Russia}
\affiliation{Instituci\'{o} Catalana de Recerca i Estudis Avan\c{c}ats (ICREA) and Institut de F\'{i}sica d'Altes Energies (IFAE), Barcelona, Spain}
\affiliation{Uppsala University, Uppsala, Sweden}
\affiliation{Taras Shevchenko National University of Kyiv, Kiev, Ukraine}
\affiliation{Lancaster University, Lancaster LA1 4YB, United Kingdom}
\affiliation{Imperial College London, London SW7 2AZ, United Kingdom}
\affiliation{The University of Manchester, Manchester M13 9PL, United Kingdom}
\affiliation{University of Arizona, Tucson, Arizona 85721, USA}
\affiliation{University of California Riverside, Riverside, California 92521, USA}
\affiliation{Florida State University, Tallahassee, Florida 32306, USA}
\affiliation{Fermi National Accelerator Laboratory, Batavia, Illinois 60510, USA}
\affiliation{University of Illinois at Chicago, Chicago, Illinois 60607, USA}
\affiliation{Northern Illinois University, DeKalb, Illinois 60115, USA}
\affiliation{Northwestern University, Evanston, Illinois 60208, USA}
\affiliation{Indiana University, Bloomington, Indiana 47405, USA}
\affiliation{Purdue University Calumet, Hammond, Indiana 46323, USA}
\affiliation{University of Notre Dame, Notre Dame, Indiana 46556, USA}
\affiliation{Iowa State University, Ames, Iowa 50011, USA}
\affiliation{University of Kansas, Lawrence, Kansas 66045, USA}
\affiliation{Louisiana Tech University, Ruston, Louisiana 71272, USA}
\affiliation{Northeastern University, Boston, Massachusetts 02115, USA}
\affiliation{University of Michigan, Ann Arbor, Michigan 48109, USA}
\affiliation{Michigan State University, East Lansing, Michigan 48824, USA}
\affiliation{University of Mississippi, University, Mississippi 38677, USA}
\affiliation{University of Nebraska, Lincoln, Nebraska 68588, USA}
\affiliation{Rutgers University, Piscataway, New Jersey 08855, USA}
\affiliation{Princeton University, Princeton, New Jersey 08544, USA}
\affiliation{State University of New York, Buffalo, New York 14260, USA}
\affiliation{University of Rochester, Rochester, New York 14627, USA}
\affiliation{State University of New York, Stony Brook, New York 11794, USA}
\affiliation{Brookhaven National Laboratory, Upton, New York 11973, USA}
\affiliation{Langston University, Langston, Oklahoma 73050, USA}
\affiliation{University of Oklahoma, Norman, Oklahoma 73019, USA}
\affiliation{Oklahoma State University, Stillwater, Oklahoma 74078, USA}
\affiliation{Brown University, Providence, Rhode Island 02912, USA}
\affiliation{University of Texas, Arlington, Texas 76019, USA}
\affiliation{Southern Methodist University, Dallas, Texas 75275, USA}
\affiliation{Rice University, Houston, Texas 77005, USA}
\affiliation{University of Virginia, Charlottesville, Virginia 22904, USA}
\affiliation{University of Washington, Seattle, Washington 98195, USA}
\author{V.M.~Abazov} \affiliation{Joint Institute for Nuclear Research, Dubna, Russia}
\author{B.~Abbott} \affiliation{University of Oklahoma, Norman, Oklahoma 73019, USA}
\author{B.S.~Acharya} \affiliation{Tata Institute of Fundamental Research, Mumbai, India}
\author{M.~Adams} \affiliation{University of Illinois at Chicago, Chicago, Illinois 60607, USA}
\author{T.~Adams} \affiliation{Florida State University, Tallahassee, Florida 32306, USA}
\author{J.P.~Agnew} \affiliation{The University of Manchester, Manchester M13 9PL, United Kingdom}
\author{G.D.~Alexeev} \affiliation{Joint Institute for Nuclear Research, Dubna, Russia}
\author{G.~Alkhazov} \affiliation{Petersburg Nuclear Physics Institute, St. Petersburg, Russia}
\author{A.~Alton$^{a}$} \affiliation{University of Michigan, Ann Arbor, Michigan 48109, USA}
\author{A.~Askew} \affiliation{Florida State University, Tallahassee, Florida 32306, USA}
% ...
%
%
\collaboration{The D0 Collaboration\footnote{with visitors from
$^{a}$Augustana College, Sioux Falls, SD, USA,
$^{b}$The University of Liverpool, Liverpool, UK,
$^{c}$DESY, Hamburg, Germany,
$^{d}$Universidad Michoacana de San Nicolas de Hidalgo, Morelia, Mexico
$^{e}$SLAC, Menlo Park, CA, USA,
$^{f}$University College London, London, UK,
$^{g}$Centro de Investigacion en Computacion - IPN, Mexico City, Mexico,
$^{h}$Universidade Estadual Paulista, S\~ao Paulo, Brazil,
$^{i}$Karlsruher Institut f\"ur Technologie (KIT) - Steinbuch Centre for Computing (SCC),
D-76128 Karlsruhe, Germany,
$^{j}$Office of Science, U.S. Department of Energy, Washington, D.C. 20585, USA,
$^{k}$American Association for the Advancement of Science, Washington, D.C. 20005, USA,
$^{l}$Kiev Institute for Nuclear Research, Kiev, Ukraine
and
$^{m}$University of Maryland, College Park, Maryland 20742, USA.
}} \noaffiliation
\vskip 0.25cm
\date{May 7, 2014}

\begin{abstract}
%
%
We measure the mass of the top quark in lepton$+$jets final states using the full sample of $p\bar{p}$ collision data collected by the D0 experiment in Run~II of the Fermilab Tevatron Collider at \mbox{$\sqrt s=1.96~$TeV}, corresponding to $9.7~\fb$ of integrated luminosity. We use a matrix element technique that calculates the probabilities for each event to result from $\ttbar$ production or background. The overall jet energy scale is constrained {\it in situ} by the mass of the $W$ boson. We measure 
$m_t=174.98\pm0.76~\GeV$. 
This constitutes the most precise single measurement of the top-quark mass.
\end{abstract}
\maketitle


Since its discovery~\cite{bib:discoverydzero,bib:discoverycdf}, the determination of the properties of the top quark has been one of the main goals of the Fermilab Tevatron Collider, recently joined by the CERN Large Hadron Collider. The measurement of the top quark mass \mt, a fundamental parameter of the standard model (SM), has received particular attention.
Indeed, \mt, the mass of the $W$ boson \mw, and the mass of the Higgs boson are related through radiative corrections that provide an internal consistency check of the SM~\cite{bib:lepewwg}. Furthermore, \mt dominantly affects the stability of the SM Higgs potential, which has related cosmological implications~\cite{bib:vstab1,bib:vstab2,bib:vstab3}.
Currently, with $\mt=173.34\pm0.76~\GeV$, a world-average combined precision of about 0.5\% has been achieved~\cite{bib:combitevprd,bib:combitev,bib:combiworld}.

In this Letter, we present a measurement of \mt using 
a matrix element (ME) technique, which determines the probability of observing each event under both the $\ttbar$ signal and background hypotheses described by the respective MEs~\cite{bib:run1nature}.
The overall jet energy scale (JES) is calibrated {\it in situ} by constraining the reconstructed invariant mass of the hadronically decaying $W$ boson to $\mw=80.4$~GeV~\cite{bib:wmass}. The measurement is performed using the full set of $p\bar p$ collision data at $\sqrt s=1.96~$TeV recorded by the D0 detector in the Run II of the Fermilab Tevatron Collider, corresponding to an integrated luminosity of $9.7~\fb$. This is an update of a previous D0 measurement that used 3.6~\fb of integrated luminosity and measured $\mt=174.94\pm1.14\thinspace({\rm stat+JES})\pm0.96\thinspace({\rm syst})~\GeV$~\cite{bib:mtprd}. In the present measurement, we not only use a larger data sample to improve the statistical precision, but also refine the estimation of systematic uncertainties through an updated detector calibration, in particular improvements to the $b$-quark JES corrections~\cite{bib:jes}, and using recent improvements in  modeling the \ttbar signal. The analysis was performed blinded in \mt.

The D0 detector central-tracking system consists of a 
silicon microstrip tracker and a central fiber tracker, 
both located within a 1.9~T superconducting solenoidal 
magnet~\cite{run2det,run2lyr0}, with designs optimized for tracking and 
vertexing at pseudorapidities $|\eta|<3$ and $|\eta|<2.5$, respectively~\cite{bib:coor}.
A liquid-argon calorimeter with uranium absorber plates has a 
central section covering pseudorapidities up to 
$|\eta| \approx 1.1$, and two end calorimeters that extend coverage 
to $|\eta|\approx 4.2$, with all three housed in separate 
cryostats~\cite{run1det}. An outer muon system, at $|\eta|<2$, 
consists of a layer of tracking detectors and scintillation trigger 
counters in front of 1.8~T iron toroids, followed by two similar layers 
after the toroids~\cite{run2muon}. 

The top quark decays into a $b$ quark and a $W$ boson with $\approx100\%$ probability assuming unitarity of the CKM matrix, resulting in a $W^+W^-b\bar b$ final state. This analysis is performed using lepton$+$jets (\ljets) final states, where one of the $W$ bosons decays leptonically, and the other hadronically. Here, $\ell$ denotes either an electron~($e$) or a muon~($\mu$), including those from leptonic tau decays. This analysis requires the presence of one isolated electron~\cite{bib:emid} or muon~\cite{bib:muid} with  transverse momentum $\pt>20~\GeV$ and $|\eta|<1.1$ or $|\eta|<2$, respectively. In addition, exactly four jets with $\pt>20~\GeV$ within $|\eta|<2.5$, and $\pt>40~\GeV$ for the jet of highest \pt, are required.  Jets are reconstructed using an iterative cone algorithm~\cite{bib:cone} with a cone parameter of $R=0.5$. Jet energies are corrected to the particle level using calibrations derived from exclusive $\gamma+$jet, $Z+$jet, and dijet events~\cite{bib:jes}. These calibrations account for differences in detector response to jets originating from a gluon, a $b$~quark, and $u,d,s,$ or $c$~quarks. 
Furthermore, each event must have an imbalance in transverse momentum of $\met>20~\GeV$ expected from the undetected neutrino.
Additional selection requirements to suppress background contributions from multijet (MJ) production are discussed in more detail in Ref.~\cite{bib:diffxsec}.
To further reduce background, at least one jet per event is required to be tagged as originating from a $b$ quark ($b$-tagged) through the use of a multivariate algorithm~\cite{bib:bid}. The tagging efficiency is on average $\approx65\%$ for $b$-quark jets in this analysis, while the mistag rate for gluons and for light ($u,d,s$) quark jets is $\approx5\%$. In total, 1468 and 1124 events are selected in the \ejets and \mujets channels, respectively, which is consistent with expectation from SM predictions.

The extraction of \mt is based on the kinematic information in the event and performed with a likelihood technique using per-event probability densities (PD) defined by the MEs of the processes contributing to the observed events. 
Assuming only two non-interfering contributing processes, \ttbar and $W+{\rm jets}$ production, the per-event PD is:
\begin{eqnarray}
\pevt &=& A(\x)[ f\psig(\x; \mt,\kjes)\nonumber\\
      &+& (1-f)\pbkg(\x;\kjes) ]\,,\label{eq:pevt}
\end{eqnarray}
where the observed signal fraction $f$, \mt, and the overall multiplicative factor adjusting the energies of jets after the JES calibration $\kjes$, are parameters to be determined from data. Here, $\x$ represents the measured jet and lepton four-momenta, and $A(\x)$ accounts for acceptance and efficiencies. 
The function \psig describes the PD for \ttbar production. Similarly, \pbkg describes the PD for $W+{\rm jets}$ production, which contributes 14\% of the data in the \ejets and 20\% in the \mujets channels according to the normalization procedure in Ref.~\cite{bib:diffxsec}. 
$W+{\rm jets}$ and MJ backgrounds have similar PD in the studied kinematic region, and thus MJ production is accounted for in \pevt via \pbkg. MJ events contribute 12\% to the \ejets and 5\% to the \mujets channels. The combined contribution from all other backgrounds amounts to about 5\% in both channels.

In general, the set $\x$ of measured quantities will not be identical to the set of corresponding partonic variables $\y$ because of finite detector resolution and parton hadronization. Their relationship is described by the transfer function $W(\x,\y,\kjes)$, where we assume that the jet and lepton angles are known perfectly. The densities \psig and \pbkg are calculated through a convolution of the differential partonic cross section, $\dif\sigma(\y)$, with $W(\x,\y,\kjes)$ and the PDs for the initial-state partons, $f(q_i)$, where the $q_i$ are the momenta of the colliding partons, by integrating over all possible parton states leading to $\x$:
\begin{eqnarray}
\psig = \frac1{\sigma_{\rm obs}^{\ttbar}(\mt,\kjes)}\int\sum&&\!\!\!\!\!\!\dif\sigma(\y,\mt)\dif \vec q_1\dif\vec q_2 f(\vec q_1)f(\vec q_2)\nonumber\\
&&\!\!\!\!\!\times\, W(\x,\y;\kjes)\,.\label{eq:psig}
\end{eqnarray}
The sum in the integrand extends over all possible flavor combinations of the initial state partons. The longitudinal momentum parton density functions (PDFs), $f(q_{i,{\rm z}})$, are taken from the CTEQ6L1 set~\cite{bib:cteq}, while the dependencies $f(q_{i,{\rm x}})$, $f(q_{i,{\rm y}})$ on transverse momenta are taken from PDs obtained from the \pythia simulation~\cite{bib:pythia}. The factor ${\sigma_{\rm obs}^{\ttbar}(\mt,\kjes)}$, defined as the expected total $\ttbar$ cross section,
ensures that $A(\x)\psig$ is normalized to unity. The differential cross section, $\dif\sigma(\y,\mt)$, is calculated using the leading order (LO) ME for the process $q\bar q\to\ttbar$. The integration in Eq.~\ref{eq:psig} is performed over the masses of the $t$ and $\bar t$ quarks which are assumed to be equal, the masses of the $W^\pm$ bosons, the energy $E$ (curvature $1/\pt$) of the electron (muon), and $E_q/(E_q+E_{\bar q})$ for the quarks from the $W\to q\bar q'$ decay. The $M_W = 80.4$~GeV constraint for the {\it in-situ} JES calibration is imposed by integrating over $W$ boson masses from a Breit-Wigner prior. There are 24 possible jet-parton assignments that are summed with weights based on their consistency with the $b$-tagging information.

The density \psig is calculated by numerical Monte Carlo (MC) integration and is identical to that in Ref.~\cite{bib:mtprd}, except as described. The transfer function $W(\x,\y;\kjes)$ and $\sigma_{\rm obs}^{\ttbar}(\mt,\kjes)$ are rederived using improved detector calibrations. Instead of pseudo-random numbers, we utilize the implementation of Bratley and Fox~\cite{bib:bratley} of the Sobol low discrepancy sequence~\cite{bib:sobol} for MC integration, which provides a reduction of about one order of magnitude in calculation time. Furthermore, we approximate the exact results of Eq.~(\ref{eq:psig}) for a grid of points in $(\mt,\kjes)$ space by calculating the ME only once for each $\mt$ and multiplying the results with the transfer function $W(\x,\y;\kjes)$ to obtain \psig for any \kjes. This results in another order of magnitude reduction in computation time. Both improvements are verified to provide a performance of the ME technique consistent with that in Ref.~\cite{bib:mtprd}. They proved essential to reduce the statistical uncertainty in evaluating most of the systematic uncertainties discussed below.

The differential partonic cross section for \pbkg is calculated using the LO $W+4{\rm~jets}$ MEs implemented in \vecbos~\cite{bib:vecbos}. The initial-state partons are all assumed to have zero transverse momentum $\pt$. As in the case of \psig, we apply identical procedures to calculate \pbkg to those in Ref.~\cite{bib:mtprd}, but using the updated transfer function $W(\x,\y;\kjes)$ and background normalization factor.

We calculate \psig and \pbkg on a grid in $(\mt,\kjes)$ with spacings of $(1~\GeV,0.01)$. A likelihood function, ${\cal L}(\x_1,\x_2,...,\x_N;\mt,\kjes,f)$, is constructed at each grid point from the product of the individual \pevt values for the measured quantities $\x_1,\x_2,...,\x_N$ of the selected events, and $f$ is determined by maximizing $\cal L$ at that grid point. The likelihood function ${\cal L}(\x_1,\x_2,...,\x_N;\mt,\kjes)$ is then projected onto the $\mt$ and $\kjes$ axes by integrating over \kjes and \mt, respectively. Best unbiased estimates of \mt and \kjes and their statistical uncertainties are extracted from the mean and standard deviation (SD) of  ${\cal L}(\x_1,\x_2,...,\x_N;\mt)$ and  ${\cal L}(\x_1,\x_2,...,\x_N;\kjes)$.

Simulations are used to calibrate the ME technique. 
Signal \ttbar events, as well as the dominant background contribution from $W+{\rm jets}$ production, are generated with \alpgen~\cite{bib:alpgen} using the CTEQ6L1 set of PDFs, interfaced to \pythia for parton showering using the MLM matching scheme~\cite{bib:mlm}. 
Therefore, it is the value of \mt as defined in the MC generator that is measured, and this value is expected to correspond within $\approx$ 1 GeV to \mt as defined in the pole mass scheme~\cite{bib:pdg}.
The simulation of parton showers with \pythia uses modified tune~A with the CTEQ6L1 PDF set and fixed $\Lambda_{\rm QCD}$. The detector response is fully simulated through {\sc geant3}~\cite{bib:geant}, followed by the same reconstruction algorithms as used on data. See Ref.~\cite{bib:diffxsec} for more details on MC simulations. Contributions from MJ production are estimated with the ``matrix method''~\cite{bib:diffxsec} and modeled using a data sample, where lepton isolation requirements are inverted.

Seven samples of \ttbar events, five at $\mt^{\rm gen}=165, 170,$ $172.5, 175, 180~\GeV$ for $\kjes^{\rm gen}=1$, and two at $\kjes^{\rm gen}=0.95, 1.05$ for $\mt^{\rm gen}=172.5~\GeV$, are generated. 
Three samples of $W+{\rm jets}$ events, at $\kjes^{\rm gen}=0.95,1,$ and $1.05$, are produced. Together, the \ttbar, $W+{\rm jets}$ and MJ samples are used  to derive a linear calibration for the response of the ME technique to \mt and \kjes. For each generated $(\mt^{\rm gen},\kjes^{\rm gen})$ point, 1000 pseudo-experiments  (PE) are constructed, each containing the same number of events as observed in data. This is done by randomly drawing simulated signal and background events according to the signal fraction $f$ from Eq.~\ref{eq:pevt}, which is randomly varied according to a binomial distribution around the value measured in data. Each of the PEs contains the number of MJ events determined from the matrix method.

The signal fraction $f$ used to construct PEs for the calibration of the method response in \mt and \kjes is extracted from data by maximizing 
the likelihood after integrating over \mt and \kjes.
Five sets of PEs are formed, for $f=0.5,0.6,0.7,0.8,$ and $0.9$ at $\mt^{\rm gen}=172.5~\GeV,\kjes^{\rm gen}=1$ to linearly calibrate the response of the ME technique to $f$. We find $f=63\%$ in the \ejets and $f=70\%$ in the \mujets channels, with an absolute uncertainty  of 1\% due to the finite size of the data sample and the calibration in $f$. 
These values are in agreement with the expectation for the signal yield assuming $\sigma_{\ttbar}=7.24~{\rm pb}$~\cite{bib:xsec}.

\begin{figure}
\begin{centering}
\includegraphics[width=0.49\columnwidth]{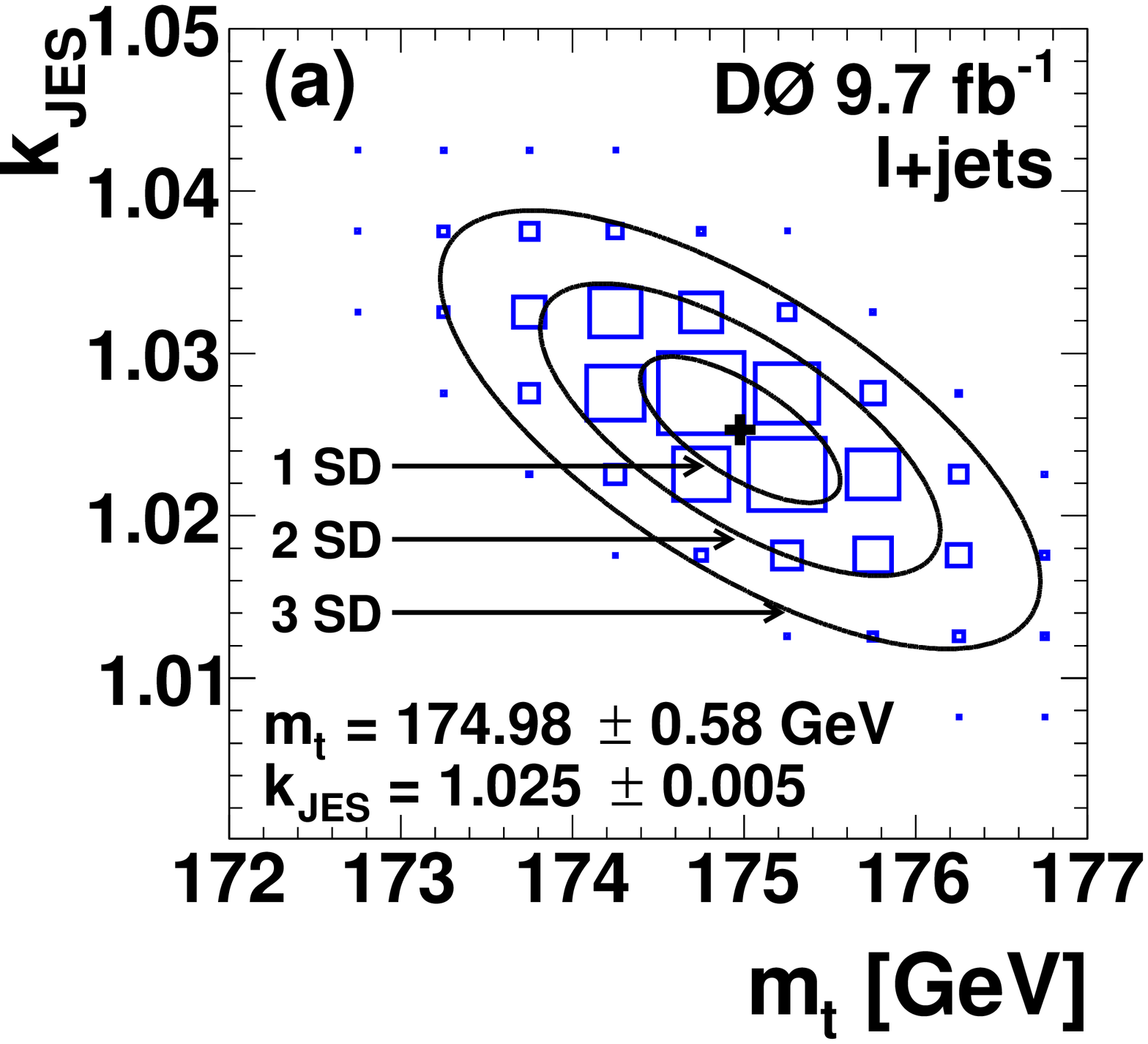}
\includegraphics[width=0.49\columnwidth]{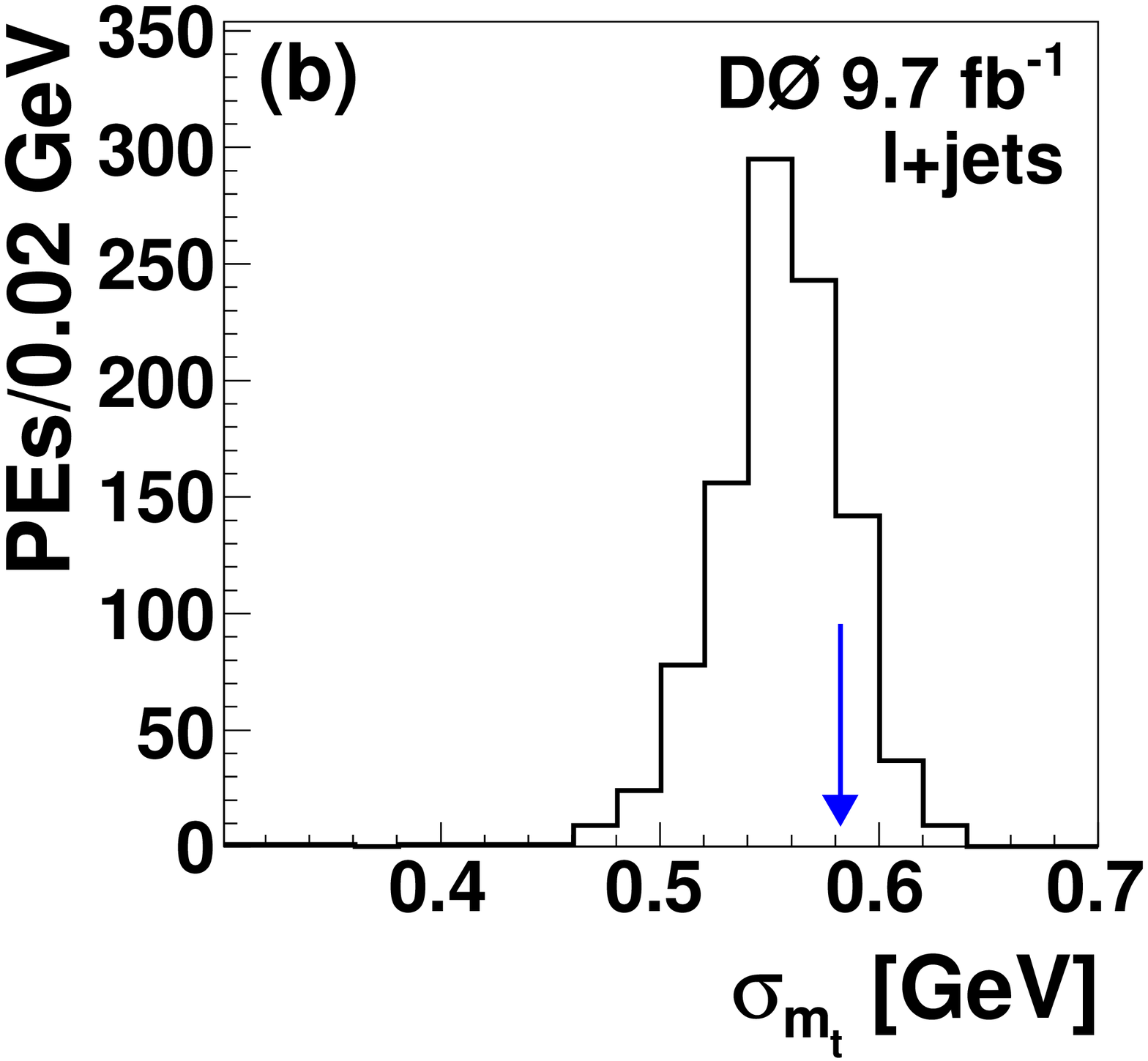}
\par\end{centering}
\caption{
\label{fig:like2d}
(color online) 
(a) Two-dimensional likelihood ${\cal L}(\x_1,\x_2,...,\x_N;\mt,\kjes)/{\cal L}_{\rm max}$ for data. Fitted contours of equal probability are overlaid as solid lines. The maximum is marked with a cross. Note that the bin boundaries do not necessarily correspond to the grid points on which $\cal L$ is calculated.
(b) Expected uncertainty distributions for \mt with the measured uncertainty indicated by the arrow.
}
\end{figure}

With $f$ determined as above, we proceed to form PEs at the chosen $(\mt^{\rm gen},\kjes^{\rm gen})$ points, and extract linear calibrations of the ME technique response to \mt and \kjes. Applying them to data, we measure 
$\mt=174.98\pm0.58~\GeV$ and $\kjes=1.025\pm0.005\,,$ 
where the total statistical uncertainty on \mt also includes the statistical contribution from \kjes. Both uncertainties are corrected by the observed SD of the pull distributions~\cite{bib:pull}. The two-dimensional likelihood distribution in $(\mt,\kjes)$ is shown in Fig.~\ref{fig:like2d}(a). Figure~\ref{fig:like2d}(b) compares the measured total statistical uncertainty on \mt with the distribution of this quantity from the PEs at $\mt^{\rm gen}=172.5~\GeV$ and $\kjes^{\rm gen}=1$. In contrast to the previous measurement~\cite{bib:mtprd}, we do not use the JES determined in exclusive $\gamma+$jet and dijet events with an uncertainty of $\approx2\%$ to constrain \kjes. 
We follow this strategy because the statistical uncertainty on the measured \kjes value is substantially smaller than the typical uncertainty on the JES, and because \kjes relates jet energies at detector level to parton energies, while JES relates jet energies at detector level to jet energies at particle level.
%
%
%
%
Splitting the total statistical uncertainty into two parts from \mt alone and \kjes, we obtain  $\mt=174.98\pm0.41\thinspace({\rm stat})\pm0.41\thinspace({\rm JES})~\GeV$.

\begin{figure}
\begin{centering}
\includegraphics[width=0.49\columnwidth]{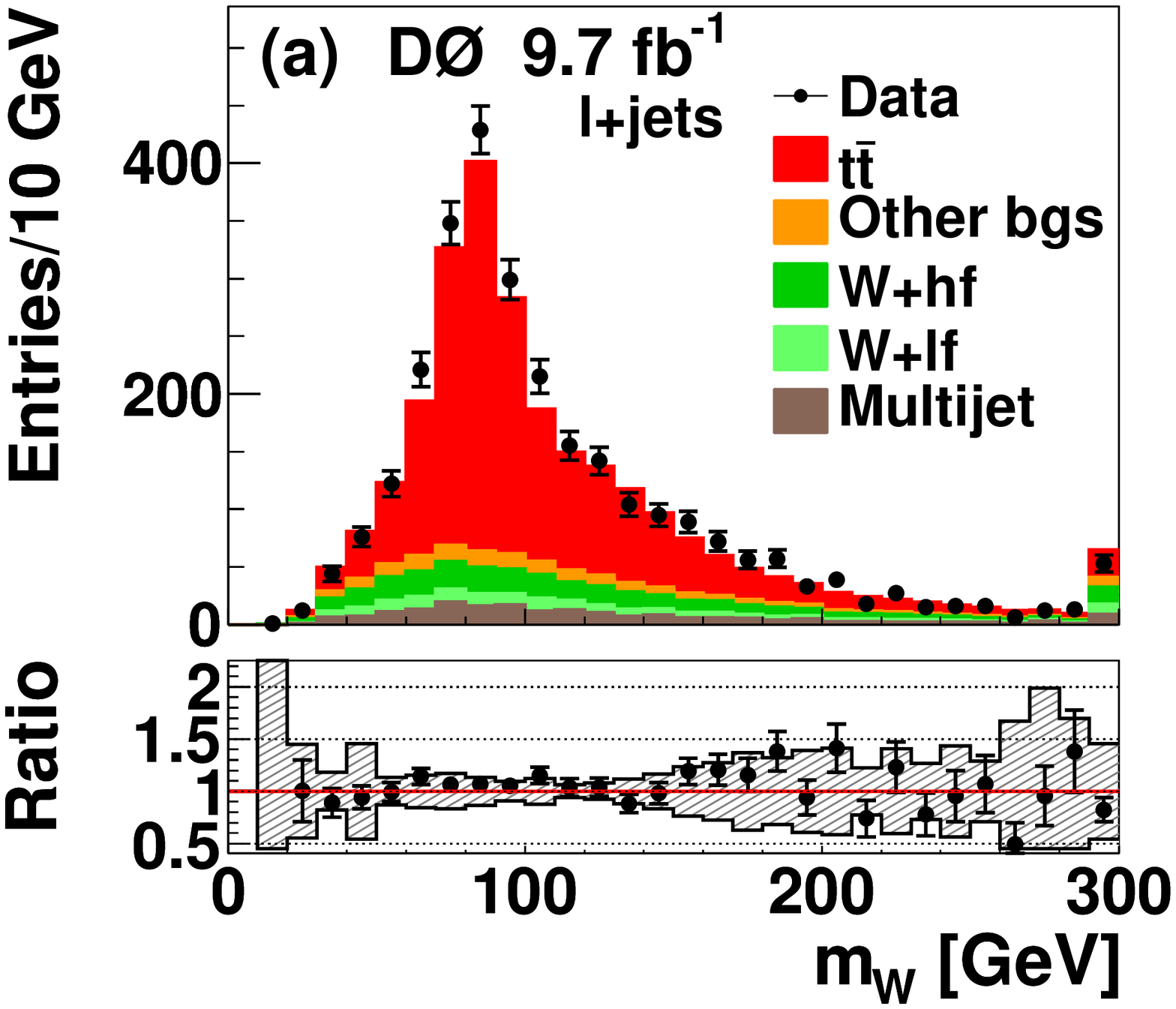}
\includegraphics[width=0.49\columnwidth]{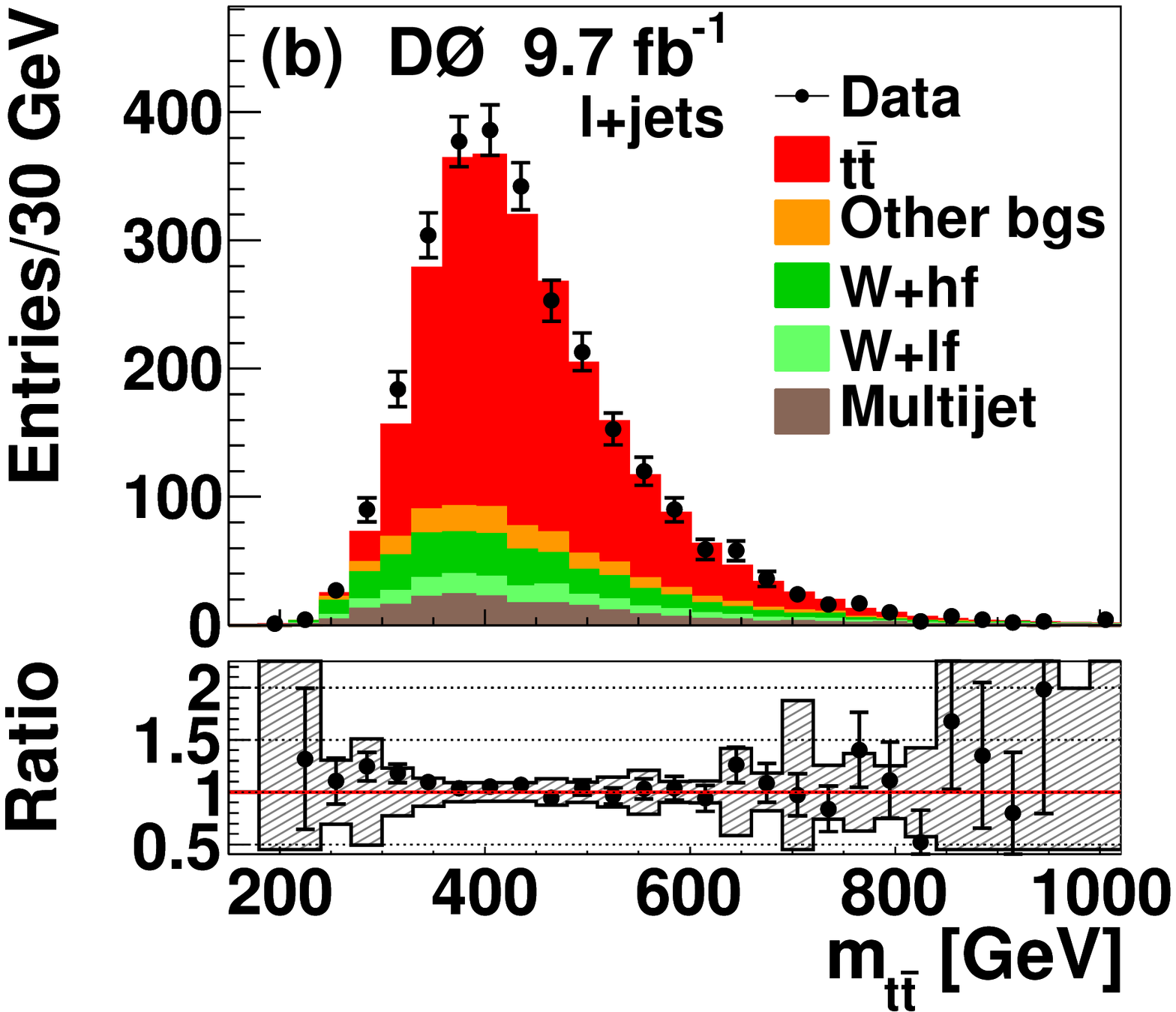}
\par\end{centering}
\caption{
\label{fig:sel}
(color online) 
(a) Invariant mass of the jet pair matched to one of the $W$ bosons.
(b) Invariant mass of the $\ttbar$ system. 
In the ratio of data to SM prediction, the total systematic uncertainty is shown as a shaded band. 
}
\end{figure}

Comparisons of SM predictions to data for $\mt=175~\GeV$ and $\kjes=1.025$ are shown in Fig.~\ref{fig:sel} for the invariant mass of the jet pair matched to one of the $W$ bosons and the invariant mass of the $\ttbar$ system.  The kinematic reconstruction is identical to the one used in Ref.~\cite{bib:diffxsec}. The \ttbar signal is normalized to total cross sections of $\sigma_{\ttbar}=7.8~$pb in the \ejets and $\sigma_{\ttbar}=7.6~$pb in the \mujets channel, corresponding to the measured signal fraction.

Systematic uncertainties are evaluated using PEs constructed from simulated signal and background events, for three categories:
modeling of signal and background events,
uncertainties in the simulation of the detector response,
and uncertainties associated with procedures used and assumptions made in the analysis. Contributions from these sources are listed in Table~\ref{tab:syst}.

The first four sources of systematic uncertainty in Table~\ref{tab:syst} are evaluated for $\mt^{\rm gen}=172.5~\GeV$ by comparing results for \mt using different signal models. 
All other systematic uncertainties are evaluated by rederiving the calibration with simulations reflecting an alternative model, and applying the alternative calibration to data.  
The statistical components of systematic uncertainties are $\approx0.05~\GeV$ for the former and $\approx0.01~\GeV$ for the latter sources of systematic uncertainty. 
The statistical components are never larger than the net difference between the default and alternative models for any of the sources of systematic uncertainty. 
One-sided sources of systematic uncertainties are taken as symmetric in both directions in the total quadrature sum.
%
\begin{table}
\begin{centering}
\begin{tabular}{lc}
\hline
\hline 
Source of uncertainty & Effect on \mt (GeV) \\
\hline
\emph{Signal and background modeling:} & \\
~~Higher order corrections & $+0.15$ \\
~~Initial/final state radiation & $\pm0.09$ \\
~~Hadronization and UE & $+0.26$ \\
~~Color reconnection & $+0.10$ \\
~~Multiple $p\bar p$ interactions & $-0.06$ \\
~~Heavy flavor scale factor & $\pm0.06$\\
~~$b$-jet modeling & $+0.09$ \\
~~PDF uncertainty & $\pm0.11$ \\
\emph{Detector modeling:} & \\
~~Residual jet energy scale & $\pm0.21$\\
~~Flavor-dependent response to jets & $\pm0.16$\\
~~$b$ tagging & $\pm0.10$\\
~~Trigger & $\pm0.01$\\
~~Lepton momentum scale & $\pm0.01$\\
~~Jet energy resolution & $\pm0.07$\\
~~Jet ID efficiency & $-0.01$\\
\emph{Method:} & \\
~~Modeling of multijet events & $+0.04$\\
~~Signal fraction & $\pm0.08$\\
~~MC calibration & $\pm0.07$\\
\hline 
{\em Total systematic uncertainty} & $\pm0.49$\\
{\em Total statistical uncertainty} & $\pm0.58$\\
{\em Total uncertainty} & $\pm0.76$\\
\hline
\hline
\end{tabular}
\par\end{centering}
\caption{
\label{tab:syst}
Summary of uncertainties on the measured top quark mass. The signs indicate the direction of the change in \mt when replacing the default by the alternative model.
}
\end{table}

We refine the evaluation procedure for several sources of systematic uncertainty compared to Ref.~\cite{bib:mtprd}  as described below. Details on other, typically smaller, sources of systematic uncertainty can be found in Ref.~\cite{bib:mtprd}. 
%
The uncertainty from {\em higher order corrections} is evaluated by comparing events simulated with {\sc mc@nlo}~\cite{bib:mcnlo} to \alpgen interfaced to \herwig~\cite{bib:herwig}.
%
The uncertainty due to the modeling of {\em initial and final state radiation} is constrained from Drell-Yan events~\cite{bib:isr}. As indicated by these studies, we change the amount of radiation via the renormalization scale parameter for the matching scale in \alpgen interfaced to \pythia~\cite{bib:isrmangano} up and down by a factor of 1.5. In addition, we reweight \ttbar simulations in $\pt$ of the \ttbar system ($\pt^{\ttbar})$ to match data, and combine the two effects in quadrature.
%
The uncertainty originating from the choice of a model for {\em hadronization and underlying event}~(UE) is evaluated by comparing events simulated with \alpgen interfaced to either \pythia or \herwig. The JES calibration is derived using \pythia with a modified tune~A~\cite{bib:jes}, and is expected to be valid for this configuration only. Applying it to events that use \herwig for evolving parton showers can lead to a sizable effect on \mt.  
However, this effect would not be present if the JES calibration were based on \herwig. To avoid such double-counting of uncertainty sources, we evaluate the uncertainty from hadronization and UE by considering as $\x$ the momenta of particle level jets matched in $(\eta,\phi)$ space to reconstructed jets. 
In this evaluation, we reweight our default \ttbar simulations in $\pt^{\ttbar}$ to match \alpgen interfaced to \herwig.
%
A potential effect of {\em color reconnection} (CR) on \mt is evaluated by comparing \alpgen events interfaced to \pythia with the Perugia 2011NOCR and Perugia 2011 tunes~\cite{bib:perugia}, where the latter includes an explicit CR model.
%
The {\em residual jet energy scale} uncertainty from a potential dependence of the JES on $(\pt,\eta)$ is estimated by changing the jet momenta as a function of $(\pt,\eta)$ by the upper limits of JES uncertainty, the lower limits of JES uncertainty, and a linear fit within the limits of JES uncertainty. The maximum excursion in \mt is quoted as systematic uncertainty.
%
Dedicated calibrations to account for the {\em flavour-dependent response to jets} 
 originating from a gluon, a $b$ quark and $u,d,c,$ or $s$ quarks
are now an integral part of the JES correction~\cite{bib:jes}, and the uncertainty on \mt from these calibrations is evaluated by changing them within their respective uncertainties. This systematic uncertainty accounts for the difference in detector response to $b$- and light-quark jets.
%
To evaluate the uncertainty from modeling of {\em $b$ tagging}, differential corrections in $(\pt,\eta)$ to ensure MC -- data $b$-tagging efficiency agreement are changed within their uncertainties.
%
The uncertainty due to the {\em modeling of multijet events} is evaluated by assuming a 100\% uncertainty on its contribution to the data sample, i.e., by leaving it out when deriving the alternative calibration.
%
We construct PEs with $\pm5\%$ variations on the measured {\em signal fraction}, which approximately corresponds to the systematic uncertainty on the measured \ttbar production cross section using D0 data~\cite{bib:xsec54}, ignoring the uncertainty from integrated luminosity, and construct the PEs according to this 5\% change.
%
%
%
%
%

In summary, we have performed a measurement of the mass of the top quark using the matrix element technique in $\ttbar$ candidate events in lepton$+$jets final states using 9.7~\fb of Run~II integrated luminosity collected by the D0 detector at the Fermilab Tevatron $p\bar p$ Collider. The result, 
\begin{eqnarray*}
\mt &=& 174.98 \pm 0.58\thinspace({\rm stat+JES}) \pm 0.49\thinspace({\rm syst})~\GeV\,,~{\rm or}\\ 
\mt &=& 174.98 \pm 0.76~\GeV\,, 
\end{eqnarray*}
is consistent with the values given by the current Tevatron and world combinations of the top quark mass~\cite{bib:combitev,bib:combiworld} and achieves by itself a similar precision. With an uncertainty of $0.43\%$, it constitutes the most precise single measurement of the top quark mass, with a total systematic uncertainty notably smaller than any other single measurement.

%
We thank the staffs at Fermilab and collaborating institutions,
and acknowledge support from the
DOE and NSF (USA);
CEA and CNRS/IN2P3 (France);
MON, NRC KI and RFBR (Russia);
CNPq, FAPERJ, FAPESP and FUNDUNESP (Brazil);
DAE and DST (India);
Colciencias (Colombia);
CONACyT (Mexico);
NRF (Korea);
FOM (The Netherlands);
STFC and the Royal Society (United Kingdom);
MSMT and GACR (Czech Republic);
BMBF and DFG (Germany);
SFI (Ireland);
The Swedish Research Council (Sweden);
and
CAS and CNSF (China).
%



\begin{thebibliography}{99}
\bibitem{bib:coor}
The D0 coordinate system is right-handed, with the $\rm z$-axis pointing in the direction of the Tevatron proton beam and the $\rm y$-axis pointing upwards. The angles $\phi$ and $\theta$ are the azimuthal and polar angles relative to the $\rm x$ and $\rm z$-axes, respectively. We also use as an angular variable the pseudorapidity defined by $\eta=-\ln[\tan(\theta/2)]$.
\bibitem{bib:pull}
The pull is defined as $\sum_i(\mt^i-\langle\mt\rangle)/\Delta\mt$, where $i=1,2,...,1000$ runs over all PEs. 
We find pull widths of 1.16 and 1.19 for \mt and \kjes.
\end{thebibliography}
\end{document}